# Shelf-stable quantum-dot light-emitting diodes with high operational performance

*Desui Chen,[1#] Dong Chen,[1#] Xingliang Dai,[1#] Zhenxing Zhang,[1] Jian Lin,[1] Yunzhou Deng,[1] Yanlei Hao,[1]*

*Ci Zhang,[2] Feng Gao,[3] Haiming Zhu,[2] and Yizheng Jin[1*]*

[1] Centre for Chemistry of High-Performance & Novel Materials, State Key Laboratory of Silicon Materials, Department of Chemistry, Zhejiang University, Hangzhou 310027, China.

[2] Centre for Chemistry of High-Performance & Novel Materials, Department of Chemistry, Zhejiang University, Hangzhou 310027, China.

[3] Department of Physics, Chemistry and Biology (IFM), Linköping University, Linköping, Sweden.

[#] These authors contributed equally to this work.

Correspondence author: Prof. Yizheng Jin (yizhengjin@zju.edu.cn)

**Abstract**

Quantum-dot light-emitting diodes (QLEDs) promise high-performance and cost-effective electroluminescent devices. However, shelf stability of QLEDs, a must for practical applications, is currently overlooked. Here we reveal that the state-of-the-art QLEDs exhibit abnormal shelf-ageing behaviours, i.e., improvements in performance (positive ageing), followed by deterioration of performance (negative ageing). Mechanism studies show that the organic acids in the encapsulation resin induce in-situ reactions, which simultaneously improve electrical conductance and minimize interfacial exciton quenching at the positive-ageing stage. Progression of the in-situ reactions results in negative ageing. Inspired by these findings, we design an electron-transporting bi-layer structure, which delivers both improved electrical conductivity and suppressed interfacial exciton quenching. This design enables shelf-stable QLEDs with high operational performance, i.e., neglectable changes of external quantum efficiency (>20.0%) and record-long operational lifetime ($T_{95}$: 5,500 hours at 1,000 cd m$^{-2}$) after storage for 180 days. Our work paves the way towards the commercialization of QLEDs.

**Introduction** (up to 500 words)

QLEDs combine stable, efficient and high colour-purity emission of quantum dots (QDs) and the advantages of cost-effective solution-based processing techniques, promising large-area electroluminescent devices ideal for display and solid-state lighting applications[1-4]. In the past decade, extensive efforts have been devoted to improving the operational performance of QLEDs[3-20], including external quantum efficiency (EQE), operational lifetime, etc. The state-of-the-art QLEDs, which adopt a hybrid structure with oxide electron-transport layers (ETLs) and polymeric hole-transport layers, simultaneously demonstrate high EQEs and long operational lifetime[3,4,9,13,20]. For example, Cao et al.[13] demonstrated red QLEDs with a maximum EQE of 15.1% and an extrapolated $T_{95}$ operational lifetime (time for the luminance drops to 95% of its initial value) of ~2,300 hours at 1,000 cd m$^{-2}$ or an extrapolated $T_{50}$ operational lifetime (time for the luminance drops to 50% of its initial value) of 2,200,000 hours at 100 cd m$^{-2}$. Our group shows that the use of electrochemically-stable ligands eliminates operando reduction reactions[20], enabling red QLEDs with a maximum EQE of 16.7% and an extrapolated $T_{95}$ operational lifetime of ~3,800 hours at 1,000 cd m$^{-2}$.

This work targets shelf stability, a must for the commercialization of QLEDs. Shelf stability is currently overlooked in the QLED field. Up to now, there is no report demonstrating QLEDs with both high operational performance and long shelf lifetime. We emphasize that the QLEDs being claimed to possess long operational lifetimes of thousands or millions of hours do not necessarily possess long shelf lifetime. In practice, the ultra-long operational lifetimes of QLEDs are estimated by extrapolating the data measured under accelerated conditions[3,4,9,13,20,21]. The actual measurement time is generally within hundreds of hours. In contrast, evaluating the shelf stability of QLEDs ideally requires months or even years of shelf storage.

Remarkably, a so-called positive ageing behaviour[12,22-24], i.e., improvements in device performance after short storage time of within several days, is generally observed for the state-of-the-art QLEDs, including the two red QLEDs with exceptionally long operational lifetime[13,20]. This phenomenon implies poor shelf stability of the QLEDs, albeit in a positive way. The origin of the positive-ageing behaviour is still under debate[12,22]. Here we show that in the long-run, the QLEDs benefiting from the positive-ageing behaviours would inevitably undergo gradual deterioration of performance, which we refer to as negative ageing. Mechanism studies on the abnormal shelf-ageing behaviours inspire us to design a bi-layered oxide ETL, enabling shelf-stable QLEDs with high operational performance.

**Results and discussion**

**Positive ageing and negative ageing of QLEDs.** We use red QLEDs with the hybrid structure (**Fig. 1a**) as a model system. The devices consist of multilayers of indium-tin-oxide (ITO) anode/poly(ethylenedioxythiophene):polystyrene sulphonate (PEDOT:PSS, ~35 nm)/poly(9,9-dioctylfluorene-co-N-(4-(3-methylpropyl))diphenylamine) (TFB, ~45 nm)/QDs (~15 nm)/$Zn_{0.9}Mg_{0.1}O$ (~60 nm)/silver cathode. CdSe/CdZnSe/ZnSeS core-shell-shell QDs (**Supplementary Fig. 1**) capping with electrochemically stable tri-n-octylphosphine ligands and possessing a photoluminescence quantum yield (PL QY) of ~90% (in solution) are used as emitters. PEDOT:PSS, TFB and $Zn_{0.9}Mg_{0.1}O$ nanocrystals (**Supplementary Fig. 2**) are used as the hole-injection layer, the hole-transporting layer and the ETL, respectively. The devices were encapsulated using an ultraviolet (UV)-curable acrylic resin (LOCTITE 3492), and stored in a nitrogen-filled glovebox. We note that the acrylic resins are one of the most widely used resins for the encapsulation of electronic and optoelectronic devices[23,25].

The red QLEDs show positive ageing within short storage time (typically within 7 days). The corresponding characteristics include the increase of EQEs and current densities, and the decrease of turn-on voltages (**Figs. 1b-d**). For a typical fresh device, the peak EQE, the current density at 4.0 V and the turn-on voltage are 14.1%, 2.8 mA cm$^{-2}$ and 2.1 V, respectively. After storage for one day, these parameters change to 18.8%, 32.1 mA cm$^{-2}$ and 1.7 V, respectively.

Elongating the storage time results in negative ageing of QLEDs (**Figs. 1b-1d**). Negative ageing is characterized by the decrease of EQEs and current densities and the increase of turn-on voltages. For example, the devices stored for 21 days exhibit a peak EQE of 17.9%, a current density of 20.5 mA cm$^{-2}$ at 4.0 V and a turn-on voltage of 1.8 V (**Fig. 1d**). Extending the storage time to 70 days, these parameters change to 13.2%, 3.2 mA cm$^{-2}$ and 2.1 V, respectively (**Fig. 1d**). Another important feature of negative ageing is the developments of non-emissive areas, i.e., dark spots. Temporal evolution of the electroluminescence images (**Inset of Fig. 1e** and **Supplementary Fig. 3**) indicates the emergence of dark spots after ~14 days of storage. The dark spots spread to ~50% of the total device area after 70 days of storage.

Negative ageing of the QLEDs is also evidenced by the gradual deterioration of the operational lifetime. We quantify the operational lifetime of QLEDs by using an empirical equation, $(L_0)^n \times T_{95} = \text{constant}$[26], where $L_0$ and n are the initial brightness and acceleration factor, respectively. The acceleration factor, n is determined to be ~1.8 by fitting the $T_{95}$ values at various $L_0$ of > 5,000 cd m$^{-2}$ (typically measured at constant driving current densities of > 25.0 mA cm$^{-2}$, **Supplementary Fig. 4**). The average $T_{95}$ operational lifetimes at an $L_0$ of 10,000 cd m$^{-2}$ for the QLEDs stored for 1 day, 21 days and 70 days are decreased from ~90 hours to ~23 hours and ~2 hours, respectively (**Fig. 1e and Supplementary Fig. 5**).

We note that a sudden failure of QLEDs is often observed in the long-term operational stability measurements of ~500 h (**Supplementary Fig. 6**). For the tests starting at $L_0$ of $< 5,000$ cd m$^{-2}$, the measured $T_{95}$ lifetime are lower than the values extrapolated based on the data measured at higher $L_0$ (**Supplementary Fig. 4**). These facts imply that the factors contributing to the negative shelf ageing of the QLEDs also impact the long-term operational stability measurements.

**Organic Acid induced positive ageing and negative ageing.** To understand the shelf ageing of the red QLEDs, we investigated unencapsulated devices stored in a glovebox. Surprisingly the current density-luminance-voltage (J-L-V) and EQE-V curves of the unencapsulated devices are nearly identical after 7 days of shelf ageing (**Fig. 2a and Supplementary Fig. 7**). The absence of shelf-ageing for the devices stored in the glovebox led us to believe that the UV-curable resin results in the abnormal ageing effects, motivating us to identify its compositions. Gas chromatography-mass spectroscopy characterizations show that the main compositions are acrylic acid, N,N-dimethylacrylamide and isobornyl acrylate with a molar ratio of 1:30:10 (**Supplementary Table 1**). In order to understand the critical component(s) in the resin causing the ageing effect, three sets of unencapsulated QLEDs were separately stored at room temperature for 1 day in a nitrogen atmosphere with a saturated vapour density of pure compound of acrylic acid, N,N-dimethylacrylamide or isobornyl acrylate (see experimental for details). The N,N-dimethylacrylamide treated devices and the isobornyl-acrylate treated devices show negligible changes (**Supplementary Fig. 8**). The acrylic-acid treated devices show positive-ageing behaviours similar to those of the devices encapsulated by the UV-curable resin (**Figs. 2b and 2c**). The peak EQEs increase from 13.9% to 19.1%, accompanied by increases of the current densities by approximately one order of magnitude and a reduction of the turn-on voltage from 2.1 V to 1.7 V.

The above control experiments suggest that the component of acrylic acid, which is a commonly used additive in the acrylic resin[27], is responsible for the changes of the red QLEDs during shelf ageing. To test the general impacts of organic acids, unencapsulated QLEDs were treated by isobutyric acid, a saturated organic acid. As shown in **Fig. 2b-d** and **Supplementary Fig. 9**, the devices show positive-ageing behaviours after being exposed to a saturated vapour density of isobutyric acid in a nitrogen atmosphere for 1 day. Extending the exposure time to 7 days leads to negative-ageing characteristics, including a decrease of the peak EQE from 19.7% to 14.3%, a reduction of the current density at 4.0 V from 36.0 mA cm$^{-2}$ to 1.4 mA cm$^{-2}$, an increase of the turn-on voltage from 1.7 V to 2.2 V and spreading of the dark-spot area (**Supplementary Fig. 9**). At this point, we conclude that the vapours of organic acids induce the abnormal shelf-ageing behaviours of the QLEDs.

**Impacts of the acid-induced in-situ reactions.** We investigate the molecular mechanisms of the organic acid-induced in-situ reactions, focusing on their correlations with the shelf-ageing behaviours of the red QLEDs. To this end, various devices or films are treated by the vapour of isobutyric acid at room temperature to mimic the in-situ reactions during the shelf ageing of the red QLEDs. Isobutyric acid is selected because it is a saturated organic acid with a relatively low saturated vapour pressure[28], offering good control of the in-situ reactions.

Our previous studies on the operation mechanisms of the red QLEDs demonstrate that exciton generation is initiated by electron injection (followed by charge-confinement-enabled hole injection)[29] and exciton recombination is largely affected by the QD/ETL interfaces[17]. Based on these understanding, we propose that two factors, increase of electron injection and suppression of interfacial exciton quenching, contribute to the positive ageing of the red QLEDs. The following experiments demonstrate that both factors can be well correlated to the organic acid-induced in-situ reactions.

First, the organic acid leads to interdiffusion of silver, the cathode metal, into the oxide ETLs. X-ray photoelectron spectroscopy (XPS) depth analyses show that placing an unencapsulated QLED in a nitrogen atmosphere with a saturated vapour density of isobutyric acid for 1 day results in broadening of the silver/oxide interfacial layer from ~14 nm to ~30 nm (**Fig. 3a**).

At this stage, the interdiffusion of silver into oxide ETLs greatly increases electron conductivity. This can be evidenced by the electrical measurements on electron-only devices of ITO/aluminium/$Zn_{0.9}Mg_{0.1}O$ nanocrystals film (150 nm)/silver. After the 1-day acid treatment, the current density at 1.0 V increases from 6.4 to 432.4 mA $cm^{-2}$ (**Fig. 3b**). This result agrees well with the increase of current densities and the reduction of turn-on voltages of the red QLEDs in the positive-ageing stage.

Second, the organic acid modifies the surfaces of the $Zn_{0.9}Mg_{0.1}O$ nanocrystals. The $Zn_{0.9}Mg_{0.1}O$ nanocrystals are synthesized by hydrolysis of zinc acetate, resulting in large densities of acetate groups and hydroxide groups on the surfaces[30,31]. The organic acids readily react with the surface hydroxyl groups. This scenario is verified by spectroscopic analyses on the $Zn_{0.9}Mg_{0.1}O$ films treated by isobutyric acid. Results show that a mild acid treatment, i.e., reacting a film of $Zn_{0.9}Mg_{0.1}O$ nanocrystals (~60 nm in thickness) with a limited amount (~$10^{-7}$ mol) of isobutyric acid (see experimental for details), causes minimal changes on the absorption of the oxide-nanocrystal film (**Supplementary Fig. 10**). Meanwhile, Fourier-transform infrared spectroscopy measurements (**Fig. 3c**) show that the reaction between the free isobutyric acid and the surface hydroxyl groups substantially increases the density of the surface carboxylate groups.

The mild acid treatment effectively suppresses exciton quenching at the QD/oxide interface (**Fig. 3d**). The pristine red QD film (a monolayer of QDs) shows a PL QY of 75% and an average PL lifetime of 13.0 ns. Depositing a layer of $Zn_{0.9}Mg_{0.1}O$ nanocrystals onto the QD monolayer causes significant interfacial exciton quenching, as characterized by a lower PL QY of 51% and a shorter average PL lifetime of 8.5 ns. A mild isobutyric-acid treatment on the QD film contacting the $Zn_{0.9}Mg_{0.1}O$ layer increases the PL QY and the average PL lifetime to 74% and 12.8 ns, respectively. Furthermore, our control experiments show that acid treatment on a pristine QD film does not cause any changes to the optical properties (**Supplementary Fig. 11**). These results allow us to conclude that the mild organic-acid treatment suppresses exciton quenching by modifying the QD/oxide interfacial interactions, instead of passivating the non-radiative recombination centres in the QD films. The suppression of interfacial exciton quenching is in line with the enhancements of device EQEs in the positive-ageing stage.

Progression of the organic acid-induced in-situ reactions causes the negative ageing of QLEDs. A cross-sectional sample of a red QLED subjected to the isobutyric-acid treatment for 7 days was analysed by aberration-corrected scanning transmission electron microscopy (STEM). High-angle annular dark-field (HAADF) observations reveal severe corrosion of the silver cathode and formation of voids at the cathode/ETL interface (**Fig. 3e**). Degradation of the silver cathode is also evidenced by the drastically increased surface roughness observed in atomic force microscopy characterizations (**Supplementary Fig. 12**). Regarding the reaction between the $Zn_{0.9}Mg_{0.1}O$ films and isobutyric acid, increasing the amounts of acid causes the gradual conversion of the conductive films of oxide nanocrystals to metal carboxylate salts (**Fig. 3f and Supplementary Fig. 10**), which are insulators. Furthermore, the by-product of this reaction is water, which is readily being detected by a trace moisture analyser (**Fig. 3g and Supplementary Fig. 13**). For an encapsulated device, the progression of this reaction would cause in-situ accumulation of water, which may further trigger other in-situ chemical reactions during device storage or in-situ electrochemical reactions during device operation[26,32-36]. Combining the consequences of degradation of the silver cathode, conversion of oxide ETLs to insulating metal carboxylates and in-situ accumulation of water would inevitably lead to negative ageing of the red QLEDs.

**O-ZnO and C-ZnO inspired by the acid-induced in-situ reactions.** The above mechanism studies indicate that realizing shelf-stable QLEDs with high operational performance should reserve the benefits of the acid-induced in-situ reactions at the positive-ageing stage, i.e., enhanced electron conduction and suppressed interfacial exciton quenching while eliminating the harm caused by the in-situ reactions at the negative-ageing stage. Both electron conduction and interfacial exciton quenching can be modulated by controlling the properties of the oxide ETLs. In this regard, we designed two types of ZnO nanocrystals, namely optical ZnO (O-ZnO), which is responsible for suppressing interfacial exciton quenching, and conductive ZnO (C-ZnO), which is responsible for improving electron conductivity.

For the rational design of O-ZnO nanocrystals, we investigated the mechanism of exciton quenching at the QD/oxide interface by combining nanosecond transient absorption (TA) (**Fig. 4a**) and time-resolved (TR) PL (**Fig. 4b**) measurements. For the CdSe-based QDs, TA signal at the band-edge position probes the conduction-band electrons in QDs with negligible contributions from the valence-band holes while TR-PL is sensitive to both electrons and holes, i.e., excitons in QDs[37,38]. Therefore, both electron-transfer and hole-transfer quenching pathways lead to faster PL decay while only electron-transfer pathways yield faster TA decay kinetics. Comparing with the pristine QD films, QD films contacting the $Zn_{0.9}Mg_{0.1}O$ layers show a longer TA kinetics but a shorter TRPL kinetics (**Figs. 4a and 4b**). The results indicate nonradiative quenching of holes but not electrons in QDs by $Zn_{0.9}Mg_{0.1}O$. The photoinduced interfacial charge-transfer process, i.e., hole transfer from QDs to the $Zn_{0.9}Mg_{0.1}O$ layers, is likely to be affected by the electronic properties of the oxide-nanocrystal films[39,40]. This fact motivates us to exploit quantum confinement effects to modulate the electronic structure of the oxide nanocrystals.

We adopt a reaction system based on the hydrolysis reactions of zinc acetate and lithium hydroxide (LiOH) for the syntheses of O-ZnO nanocrystals[41,42]. We find that adjusting the concentrations of LiOH in the reaction solution results in good size control of the as-synthesized ZnO nanocrystals in the range of ~2.1 nm to ~6.2 nm (**Supplementary Fig. 14**). The ZnO nanocrystals with the smallest average size of ~2.1 nm (**Fig. 4c**) are selected because of the pronounced quantum confinement effects. Furthermore, a magnesium-acetate ($MgAc_2$) surface-modification method (see experimental for details) is developed to inhibit the post-synthesis ripening processes, leading to colloids of O-ZnO nanocrystals that are stable for months (**Supplementary Fig. 15**). The colloid of O-ZnO nanocrystals shows a first excitonic absorption peak at ~295 nm and an intra-gap PL peak at ~480 nm (**Fig. 4d**), both of which are blue-shifted comparing with those of the colloid of $Zn_{0.9}Mg_{0.1}O$ nanocrystals. On contrary to the QD/$Zn_{0.9}Mg_{0.1}O$ sample, TA and TR-PL measurements on the QD/O-ZnO sample show identical decay kinetics to those of the pristine QD films (**Figs. 4a and 4b**), indicating that the hole-transfer channels are blocked. As a result, the interfacial exciton quenching is minimized and the sample of a monolayer of QDs contacting an O-ZnO film shows a high PLQY of 74%.

Regarding C-ZnO, we show that the conductance of the films consisting of ZnO nanocrystals is highly dependent on the sizes of the nanocrystals. Electron-only devices of ITO/aluminium/ZnO film (150 nm)/silver using oxide nanocrystals with different average sizes of 2.1 to 6.2 nm were fabricated. Electrical measurements show that the current densities can be varied by three orders of magnitude (**Supplementary Fig. 16**). Thus, the ZnO nanocrystals with the largest average size, 6.2 nm (**Fig. 4e**), which is significantly larger than that of the $Zn_{0.9}Mg_{0.1}O$ nanocrystals (~3.5 nm, **Supplementary Fig. 2**), is selected as C-ZnO. Consequently, films based on the C-ZnO nanocrystals demonstrate electrical conductance two orders of magnitude larger than that of the films based on the $Zn_{0.9}Mg_{0.1}O$ nanocrystals (**Fig. 4f**).

**Shelf-stable QLEDs with high operational performance.** We design a bi-layered ETLs comprising of O-ZnO layers and C-ZnO layers for red QLEDs. Devices with a structure (**Fig. 5a**) of ITO/PEDOT:PSS (~35 nm)/TFB (~45 nm)/CdSe/CdZnSe/ZnSeS core-shell-shell QDs (~15 nm)/O-ZnO (~15 nm)/C-ZnO (~45 nm)/Ag (~100 nm) were fabricated and stored in a nitrogen-filled glovebox without encapsulation. In this structure, the O-ZnO layer is in contact with the red QDs to minimize interfacial exciton quenching and the C-ZnO layer is in contact with the silver cathode, offering efficient electron transport.

The red QLEDs demonstrate excellent optoelectronic characteristics. The fresh device exhibits a sub-bandgap turn-on voltage of 1.7 V. At a voltage of 4.0 V, the current density reaches 101.8 mA cm$^{-2}$ (**Fig. 5b**), which are higher than that of the positive-aged devices with $Zn_{0.9}Mg_{0.1}O$ ETLs (**Fig. 1b**). The fresh device demonstrates a peak EQE of 20.1% and high EQEs of >19.0% in a wide brightness range of 100-15,000 cd m$^{-2}$ (**Fig. 5c**).

We highlight that the red QLEDs with bi-layered oxide ETLs demonstrate both outstanding shelf stability and ultra-long operational lifetime. The J-L-V and EQE-L characteristics of a device are almost identical after 180 days of shelf-ageing (**Figs. 5b and 5c**). No dark spots are observed for the shelf-aged device (Inset of **Fig. 5c and Supplementary Fig. 17**). Both a fresh QLED and a 180-day shelf-aged QLED exhibit a long $T_{95}$ operational lifetime of 52 hours at ~13,300 cd m$^{-2}$ (**Fig. 5d**). Remarkably, no sudden failure is observed for all stability measurements of the QLEDs of up to ~1,400 h (**Fig. 5e**). The data of $\log T_{95}$ shows linear dependence on $\log L_0$ in a wide range of $L_0$ ranging from ~2,000 cd m$^{-2}$ to ~13,000 cd m$^{-2}$ (**Supplementary Fig.18**). Thus, we are optimistic on using the empirical extrapolating method to estimate the operational lifetime at a $L_0$ of 1,000 cd m$^{-2}$. Statistically, the average $T_{95}$ operational lifetime at 1,000 cd m$^{-2}$ for both the fresh QLEDs and the shelf-aged (for 100 days) QLEDs is determined to be ~5,500 hours (**Fig. 5f**), surpassing the record of QLEDs.

Specifically, the red QLEDs with bi-layered ETLs were encapsulated by acids-free resin (LOCTITE 3335) and stored in an air atmosphere. The fresh devices and the shelf-aged devices (100 days) show excellent and nearly identical operational performance, i.e., a turn-on voltage of 1.7 V, a luminance of > 20,000 cd m$^{-2}$ at 4.0 V, a peak EQE of >20.0% and a $T_{95}$ operational lifetime of 57 h at ~12,800 cd m$^{-2}$ (**Supplementary Fig. 19**). The results indicate that combining our device structure of bi-layered oxide ETLs with state-of-the-art encapsulation techniques can realize air-stable devices with high operational performance, which are critical for practical applications.

Finally, we show that the strategy of bi-layer oxide ETLs is readily extended to QLEDs with other types of QD emitters. Green-emitting CdZnSeS–ZnS core-shell QDs (PL peak: 522 nm) with a high PL QY of ~90% (in solution) were used as an example. Applying the device structure with the bi-layered O-ZnO/C-ZnO ETLs leads to QLEDs with a peak EQE of ~18.0% and a $T_{50}$ operational lifetime of ~100 hours at ~10,500 cd m$^{-2}$ (**Supplementary Fig. 20**). Most importantly, the devices show negligible changes after shelf-ageing for 100 days, demonstrating outstanding shelf stability.

**Conclusions**

Our work addresses shelf stability, a critical but overlooked challenge for QLEDs. State-of-the-art QLEDs possessing high efficiency and long operational lifetime demonstrate poor shelf stability, limiting their practical applications. We find that shelf ageing, both positive ageing and negative ageing, of QLEDs is originated from acid-induced in-situ reactions. The mechanism of shelf ageing is different from those of the intensively investigated operational degradation processes of QLEDs, which are mostly associated with chemical or electrochemical reactions induced by the injected charges. Our rational design of the bi-layer oxide ETLs, which eliminates the use of acid and simultaneously offers improved electrical conductance and suppressed interfacial excitons quenching, enables shelf-stable red devices with high operational performance, including record-long operational stability.

Our work suggests that shelf stability and operational stability, which are fundamentally different from each other, should be decoupled and addressed separately for future developments of QLEDs. For example, the stability of blue QLEDs remains to be improved to meet the requirements of display applications. Thus, shelf-stable devices, instead of the devices showing the abnormal shelf-ageing behaviours, shall be used for deciphering the operational degradation mechanisms of blue QLEDs. Our work also sheds light for addressing stability issues of emerging optoelectronic and electronic devices, e.g. perovskite LEDs.

**Acknowledgements**

This work was financially supported by the National Key R&D Program of China (2016YFB0401600), the National Natural Science Foundation of China (51522209, 21975220, 91833303, 51911530155 and 91733302) and the Fundamental Research Funds for the Central Universities. F.G. is a Wallenberg Academy Fellow.

**Author contributions**

Y.J. conceived the idea and supervised the work. D.C. synthesized the O-ZnO and C-ZnO nanoparticles and carried out the device fabrication and characterizations. D.C. conducted the acid treatment experiments and the optical measurements. X.D. participated in conceiving the idea and designed the experiments. Z.Z carried out the gas chromatography-mass analysis and assisted acid treatment experiments. J.L. assisted the STEM-HAADF measurements. Y.D carried out the AFM experiments. Y. H. assisted device fabrication and characterizations. C.Z. and H.Z. assisted the TA measurements. Y.J. and X.D. wrote the first draft of the manuscript. F.G. and X.P. participated in experimental design, data analysis and provided major revisions. All authors discussed the results and commented on the manuscript.

**Competing interests**

The authors declare no competing interests.

**Additional Information**

Reprints and permissions information is available at www.nature.com/reprints. Readers are welcome to

comment to the online version of the paper. Correspondence and requests for materials should be addressed Y.J. (yizhengjin@zju.edu.cn).

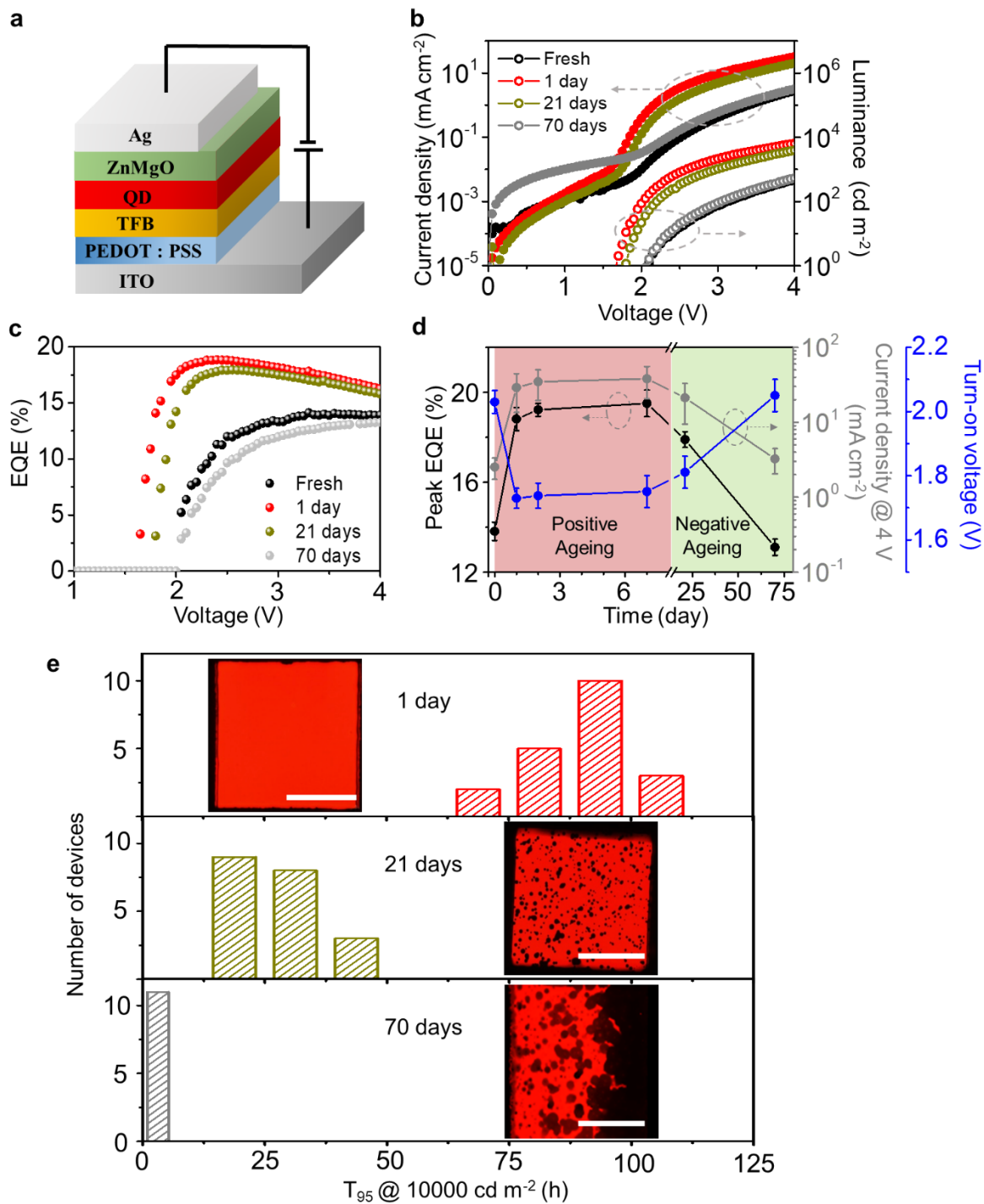

**Fig. 1 | Positive ageing and negative ageing of the red QLEDs. a,** Device structure. **b,** J-L-V

characteristics, and **c,** the corresponding EQE-V relationships for the encapsulated devices stored for 0

day (fresh device), 1 day, 21 days and 70 days in a nitrogen-filled glovebox. **d**, Shelf-ageing time-dependent peak EQEs, current densities at 4.0 V and turn-on voltages. **e**, Histograms of $T_{95}$ operational lifetime at 10,000 cd m$^{-2}$ for devices shelf-aged for 1 day (top), 21 days (middle) and 70 days (bottom), respectively. Inset: the corresponding electroluminescence images. Scale bar: 1 mm.

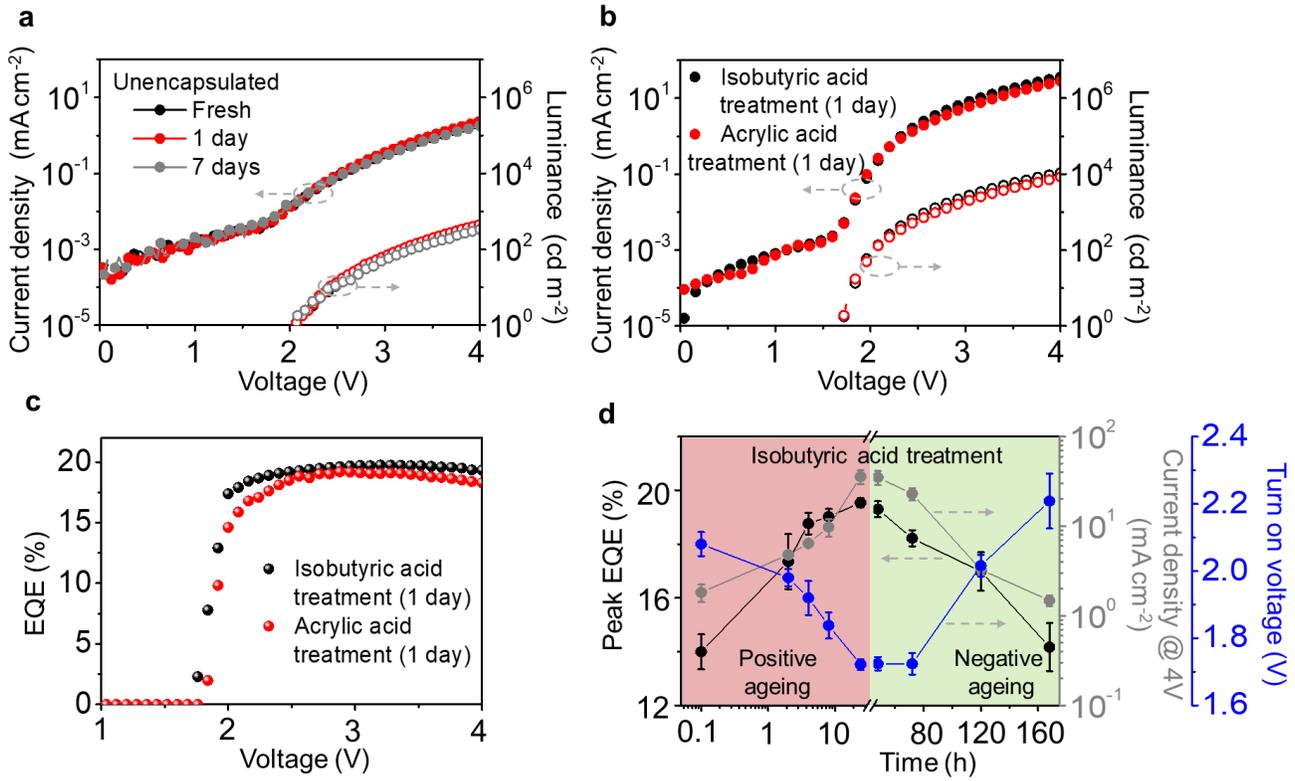

**Fig. 2 | Shelf-ageing behaviours of the unencapsulated red QLEDs and the acid-treated red QLEDs.**
**a,** J-L-V characteristics of the unencapsulated devices stored for 0 day (fresh device), 1 day and 7 days in a nitrogen-filled glovebox. **b,** J-L-V characteristics, and **c,** the corresponding EQE-V relationships for the QLEDs treated with two acids for 1 day. **d**, Dependence of peak EQEs, current densities at 4.0 V and turn-on voltages of the isobutyric-acid treated devices on shelf-ageing time.

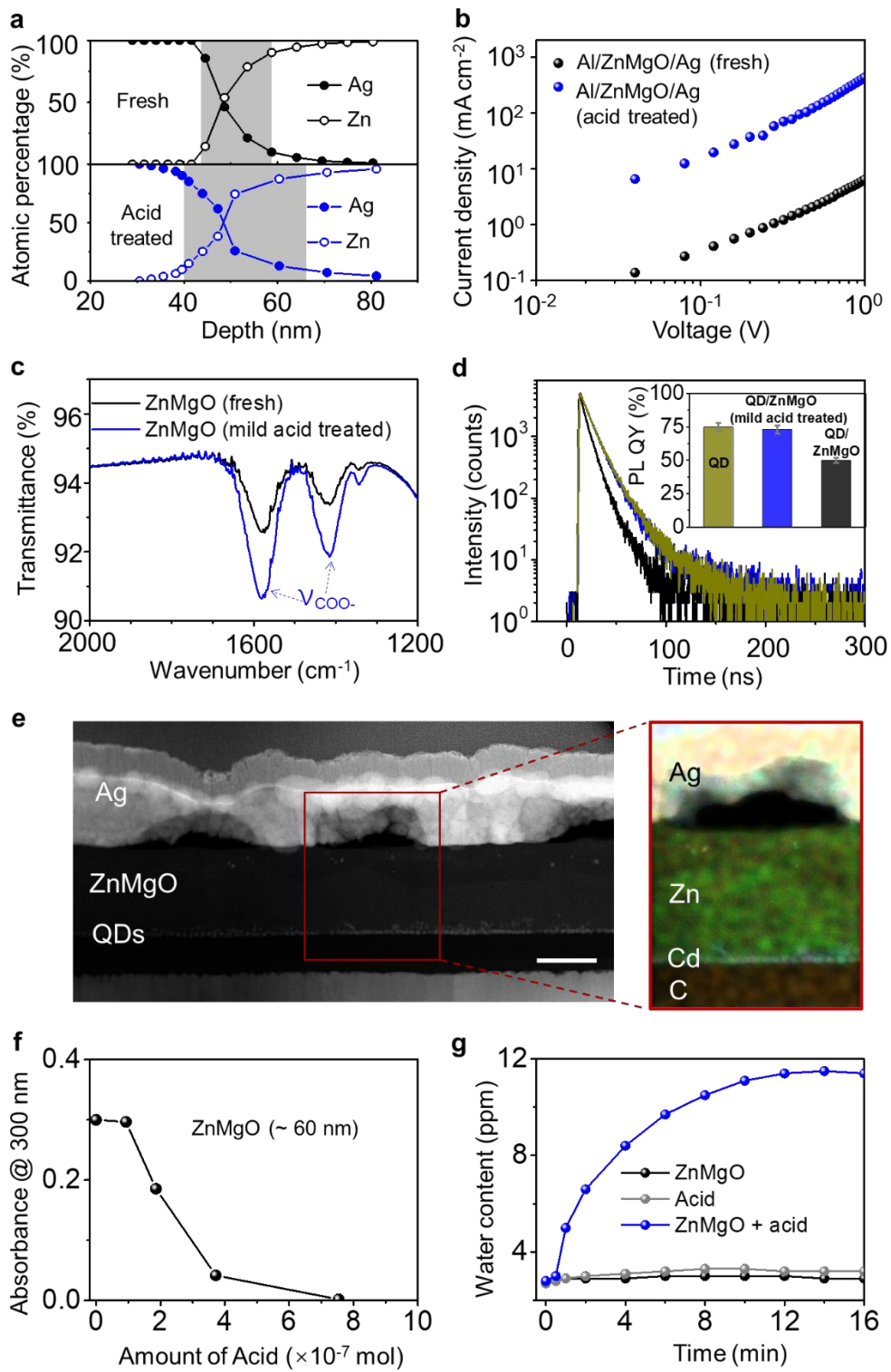

**Fig. 3 | Acid-induced in-situ reactions and their impacts. a,** XPS depth analyses showing the relative

atomic ratios of silver and zinc for the ITO/$Zn_{0.9}Mg_{0.1}O$/silver samples before (black) and after (blue) the 1-day acid treatment. The widths of the silver/$Zn_{0.9}Mg_{0.1}O$ interfacial layer are defined by the depth of ~90% (~10%) of silver (zinc) changing to ~90% (~10%) of zinc (silver). **b**, J-V curves of the electron-only devices (ITO/aluminium/$Zn_{0.9}Mg_{0.1}O$ nanocrystals (150 nm)/silver) before and after the 1-day acid treatment. **c**, Fourier-transform infrared spectra of a $Zn_{0.9}Mg_{0.1}O$ film (deposited onto a $CaF_2$ substrate) before and after the mild acid treatment. Vibration bands of the carboxylate groups are labelled as $v_{COO^-}$.

**d**, Time-resolved PL decay for the pristine QD film, the QD/$Zn_{0.9}Mg_{0.1}O$ film and the mild acid-treated QD/$Zn_{0.9}Mg_{0.1}O$ film deposited on quartz substrates. Inset showing the corresponding absolute PL QYs.

**e**, A cross-sectional STEM image of a red QLED exposed in a nitrogen atmosphere with a saturated vapour density of isobutyric acid for 7 days. Scale bar: 100 nm. Inset is a colour-mixed elemental mapping image measured by energy-dispersive X-ray spectroscopy, showing a pinhole formed at the cathode/oxide interface. **f**, Absorbance (at 300 nm, extracted from the spectra shown in **Supplementary Fig. 10**) of the $Zn_{0.9}Mg_{0.1}O$ films (deposited onto quartz substrates) reacted with different amounts of isobutyric acid. **g**, Time-dependent water content showing a significant increase of water content in 12 minutes when isobutyric acid is injected into a sealed glass bottle containing $Zn_{0.9}Mg_{0.1}O$ films (see **Supplementary Fig. 13** for detailed experimental setup). The water contents in the sealed glass bottle containing only the $Zn_{0.9}Mg_{0.1}O$ films or only the isobutyric acid are also shown.

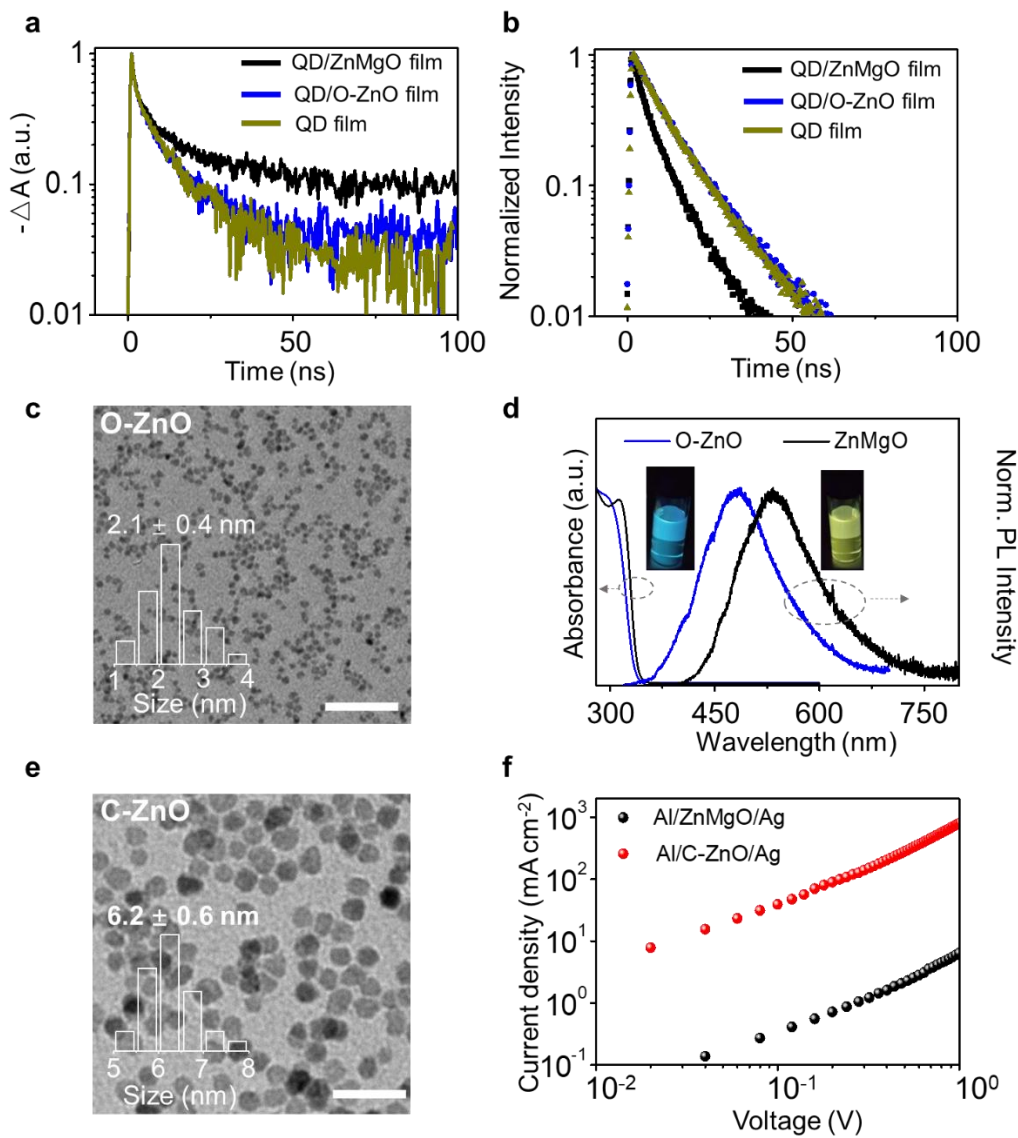

**Fig. 4 | Design of C-ZnO and O-ZnO. a,** Normalized bleaching kinetics at 2.0 eV (the first exciton absorption peak of QDs) for the QD/$Zn_{0.9}Mg_{0.1}O$ film, the QD/O-ZnO film and the QD film. **b,** The corresponding time-resolved PL decay for the QD/$Zn_{0.9}Mg_{0.1}O$ film, the QD/O-ZnO film and the QD film. **c,** A typical TEM image of the O-ZnO nanocrystals showing the sizes of 2.1 ± 0.4 nm. Scar bar: 20 nm. **d,** Absorption and PL spectra of the O-ZnO nanocrystals and the $Zn_{0.9}Mg_{0.1}O$ nanocrystals in solution. Inset showing the photograph of the colloids of the oxide nanocrystals under UV irradiation (365 nm). **e,** A typical TEM image of the C-ZnO nanocrystals showing the size of 6.2 ± 0.6 nm. Scar bar: 20 nm. **f,** J-

V curves for the electron-only devices (ITO/aluminium/oxide nanocrystals (150 nm)/silver) based on the C-ZnO nanocrystals or the $Zn_{0.9}Mg_{0.1}O$ nanocrystals.

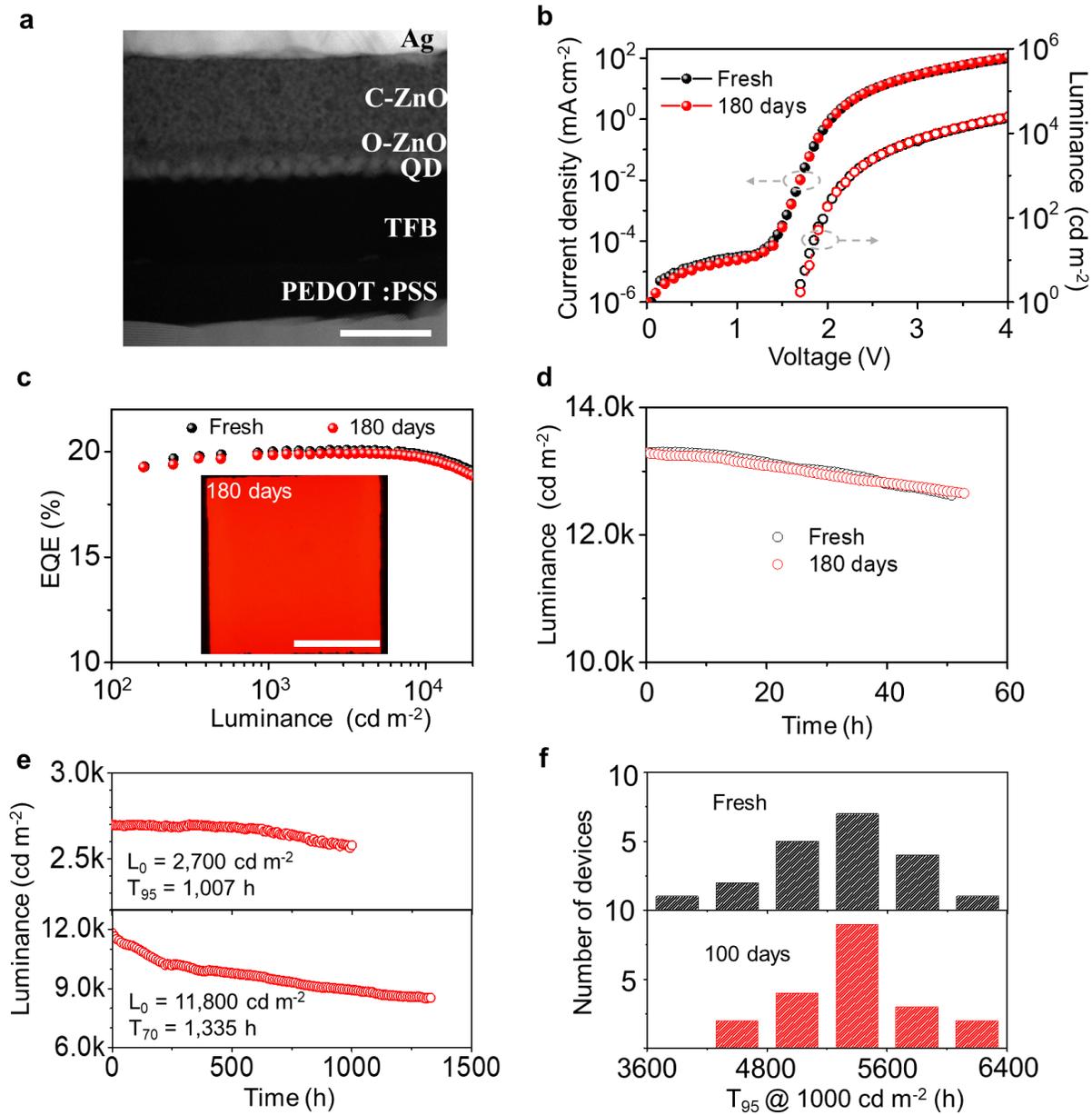

**Fig. 5 | Shelf-stable red QLEDs with bi-layered oxide ETLs. a,** A STEM-HAADF image of a cross-sectional sample showing the device structure. Scale bar: 50 nm. **b,** J-L-V characteristics, and **c,** the corresponding EQE-L relationships of a fresh device and a device shelf-aged for 180 days. Inset of **c,** an electroluminescence image of a device shelf-aged for 180 days. Scale bar: 1 mm. **d,** Stability data of a fresh device and a shelf-aged device (180 days) driven at a constant-current mode. **e,** Long-time (> 1,000 h) stability data of the two QLEDs measured at a L0 of 2,700 cd m⁻² and 11,800 cd m⁻², respectively. **f,**

Histograms of $T_{95}$ operational lifetime at 1,000 cd m$^{-2}$ measured from 20 fresh devices and 20 shelf-aged devices (100 days).

**Methods**

**Materials.** The CdSe/CdZnSe/ZnSeS core-shell-shell red QDs and the CdZnSeS-ZnS core-shell green QDs were purchased from Najing technology Co., Ltd. TFB (average molecular weight: 50,000 g mol$^{-1}$) was purchased from American Dye Source. The UV-curable resin (LOCTITE 3492 and LOCTITE 3335) were purchased from Henkel AG & Company, KGaA. Acrylic acid (99%, anhydrous, contains 200 p.p.m. MEHQ as inhibitor), isobutyric acid (99%, dry with 3A molecular sieve), tetramethylammonium hydroxide (TMAH, 98%), zinc acetate dihydrate (98%), magnesium acetate (98%) and lithium hydroxide (LiOH, 98%) were purchased from Sigma-Aldrich. Ethanol (extra dry, 99.5%) and ethyl acetate (HPLC grade) were purchased from Acros. Dimethyl sulfoxide (DMSO, HPLC grade) was purchased from Alfa-Aesar.

**Synthesis of the Zn$_{0.9}$Mg$_{0.1}$O nanoparticles.** The Zn$_{0.9}$Mg$_{0.1}$O nanoparticles were prepared by a solution-precipitation process reported in the literature with some modifications[43]. Typically, a solution of Zn(CH$_3$COO)$_2$ (2.7 mmol) and Mg(CH$_3$COO)$_2$ (0.3 mmol) in DMSO (30 mL) were mixed with a solution of TMAH (5 mmol) in ethanol (10 mL). The mixture was stirred for 1 hour under ambient conditions.

Zn$_{0.9}$Mg$_{0.1}$O nanoparticles were precipitated by adding ethyl acetate and further purified by dispersing/precipitating twice using the combination of ethanol/ethyl acetate. The resultant oxide nanocrystals were re-dissolved in ethanol and filtered (0.22 µm PTFE filter) before use.

**Device fabrication.** PEDOT:PSS solutions (Baytron PVP Al 4083, filtered through a 0.22 µm N66 filter) were spin-coated onto ITO-coated glass substrates at 3,500 r.p.m. for 40 s and baked at 150°C for 30 min in air. The PEDOT:PSS-coated substrates were treated by oxygen plasma for 4 min before transferred into a nitrogen-filled glovebox (O$_2$ < 1 p.p.m., H$_2$O < 1 p.p.m.). TFB (in chlorobenzene, 12 mg mL$^{-1}$), QDs (in octane, 15 mg mL$^{-1}$) and Zn$_{0.9}$Mg$_{0.1}$O nanoparticles (in ethanol, 30 mg mL$^{-1}$) were layer-by-layer deposited by spin coating at 2,000 r.p.m. for 45 s. The TFB layers were baked at 150℃ for 30 min before the deposition of the QD layers. For the QLEDs with bi-layered O-ZnO/C-ZnO ETLs, the O-ZnO nanoparticles (in ethanol, 8 mg mL$^{-1}$) and the C-ZnO nanoparticles (in ethanol, 20 mg mL$^{-1}$) were layer-by-layer deposited onto the QD layers. The O-ZnO layers were baked at 80℃ for 30 min before the deposition of the C-ZnO layers. Finally, silver electrodes (100 nm) were deposited through a shadow mask using a thermal evaporation system (Trovato 300C) under a base pressure of ~3× 10$^{-7}$ Torr. The device area defined by the overlapping of the ITO and silver electrodes is 4 mm$^2$. UV-curable resins (LOCTITE 3492 or LOCTITE 3335) were used to encapsulate the devices by covering glass slides in a glovebox.

**Acid treatment on the unencapsulated QLEDs or the electron-only devices.** The devices were placed onto a rack and then sealed in a glass bottle (100 mL) containing pure acid liquid (0.5 mL) for a fixed period. Treatment of the devices with N,N-dimethylacrylamide and isobornyl acrylate followed the same procedure, except the acid is replaced by N,N-dimethylacrylamide and isobornyl acrylate liquid. All procedures were conducted at room temperature in a nitrogen-filled glovebox.

**Mild Acid treatment on the $Zn_{0.9}Mg_{0.1}O$ films.** Films of $Zn_{0.9}Mg_{0.1}O$ nanocrystals or QD/$Zn_{0.9}Mg_{0.1}O$ nanocrystals were sealed in a glass bottle (20 mL). Nitrogen with a saturated vapour of isobutyric acid (1 mL, equals to ~ 10$^{-7}$ mol at 298 K) was injected into the glass bottle. The oxide films were reacted with the acid vapour for 2 hours.

**Characterizations.** The absorption spectra were obtained using an Agilent Cary 5000 spectrophotometer.

Fourier-transform infrared spectra were recorded on a Thermo IS-50 spectrophotometer.

PL spectra were measured by an Edinburgh Instruments FLS920 spectrometer.

Time-resolved PL spectra were obtained by the time-correlated single-photon counting method using an Edinburgh Instruments FLS920 spectrometer. Pulsed laser diodes (EPL-405) with a wavelength of 404.2 nm and a pulse width of 58.6 ps was used as the excitation source.

The absolute PL QYs of the QD samples were measured by applying a three-step method[44]. A system consisting of a Xenon lamp, optical fibre, a QE65000 spectrometer (Ocean Optics) and a home-designed integrating sphere was used. The samples for interfacial exciton quenching experiments were prepared by spin coating a monolayer of QDs onto a quartz substrate, followed by deposition of a layer of oxide nanocrystals (60 nm). The samples were encapsulated by using acid-free resin (LOCTITE 3335) in a nitrogen-filled glovebox before characterizations.

TA measurements were conducted on a nanosecond spectrometer system (ns-TA100, Time-Tech Spectra, LLC) with a white light supercontinuum laser as probe pulse. Transient data were collected with 400 nm excitation. To enhance the signal-to-noise ratio, samples of 20 repeated layers of QDs (a monolayer)/oxide nanocrystals (20 nm) were prepared by spin coating. Each layer was baked at 80°C for 30 min before deposition of the next layer.

Electrical characterizations of devices were performed in a nitrogen-filled glovebox ($O_2 < 1$ p.p.m., $H_2O < 1$ p.p.m. and acid-free) at room temperature. A system consisting of a Keithley 2400 source meter and an integration sphere (FOIS-1) coupled with a QE-Pro spectrometer (Ocean Optics) was used to obtain the J-L-V curves[3].

The $T_{95}$ or $T_{50}$ operational lifetimes of the QLEDs were measured by using an ageing system with an embedded photodiode designed by Guangzhou New Vision Opto-Electronic Technology Co., Ltd. All devices were bias-stressed for at least 5 hours at a constant current of 50.0 mA cm$^{-2}$ before the test.

The electroluminescence images of the working QLEDs were collected using an optical microscope (Zeiss Axio Vert).

XPS analyses were performed on a Kratos AXIS Supra spectrometer in an ultrahigh vacuum chamber with a vacuum $< 10 \times 10^{-10}$ Torr. An Al kα radiation source (1,466 eV), with an experimental resolution of 0.1 eV, was used.

The cross-sectional samples were prepared by using a dual-beam focused-ion-beam system (Quanta 3D FEG). A Cs aberration-corrected Titan G2 80-200 ChemiSTEM microscope operated at 200 kV was used for characterization.

TEM images of the nanocrystals were collected on an HT7700 microscope operated at 100 keV.

Atomic force microscopy measurements were conducted by a Cypher-S microscope placed in a nitrogen-filled glovebox.

Gas chromatography-mass analysis was performed on an Agilent 6890-5973GCMS. The sample was prepared by mixing UV-curable glue (1 mL) with dichloromethane (9 mL).

The thicknesses of the films were measured using a KLA Tencor P-7 Stylus Profiler.